\documentclass[iop]{emulateapj}
\def\axpu0142{4U~0142+61}
\def\ae1048{1E~1048.1-5937}
\def\ce1841{1E~1841-045}
\def\de2259{1E~2259+586}

\begin{document}

\shorttitle{Broadband X-ray Spectra of Magnetars}
\title{Broadband X-ray Spectral Investigations of Magnetars, \axpu0142, \ce1841,
\de2259, and \ae1048}


\author{Shan-Shan Weng$^{1}$ \& Ersin G\"o\u{g}\"u\c{s}$^{2}$}

\affil{$^{1}$ Department of Physics and Institute of Theoretical Physics,
Nanjing Normal University, Nanjing 210023, China}

\affil{$^{2}$ Sabanc\i~University, Faculty of Engineering and Natural
  Sciences, Orhanl\i ~Tuzla 34956 Istanbul Turkey}

\email{wengss@ihep.ac.cn}
\begin{abstract}

We have generated an extended version of rather simplified but physically
oriented three-dimensional magnetar emission model, STEMS3D, to allow spectral
investigations up to 100 keV. We have then applied it to the broadband spectral
spectra of four magnetars: 4U~0142+61, 1E~1841-045, 1E~2259+586 and
1E~1048.1-5937, using data collected with {\it Swift}/XRT or {\it XMM-Newton}
in soft X-rays, and {\it Nuclear Spectroscopic Telescope Array} in the hard
X-ray band. We found that the hard X-ray emission of 4U~0142+61 was spectrally
hard compared to the earlier detections, indicating that the source was likely
in a transition to or from a harder state. We find that the surface properties
of the four magnetars are consistent with what we have obtained using only the
soft X-ray data with STEMS3D, implying that our physically motivated magnetar
emission model is a robust tool. Based on our broadband spectral
investigations, we conclude that resonant scattering of the surface photons in
the magnetosphere alone cannot account for the hard X-ray emission in
magnetars; therefore, an additional non-thermal process, or a population of
relativistic electrons is required. We also discuss the implication of the
non-detection of persistent hard X-ray emission in 1E~1048.1-5937.

\end{abstract}


\section{Introduction}

Two small populations of young, isolated neutron stars (NSs), anomalous X-ray
pulsars (AXPs) and soft gamma-ray repeaters (SGRs) are considered to be
ultra-magnetized objects, $B \geq 10^{14}$ G, also known as magnetars, as is
primarily indicated by their emission of energetic bursts and their high
spin-down rates \citep{thompson95,thompson96}. To date, 28 magnetars or
magnetar candidates have been identified in the Galaxy, LMC, and SMC
\citep{olausen14} \footnote{Note that PSR J1846-0258 is usually referred to a
rotation-powered pulsar, but it also exhibited a magnetar-like burst in 2006
\citep{gavriil08}.
\protect\url{http://www.physics.mcgill.ca/$\sim$pulsar/magnetar/main.html}}.
Magnetars are luminous in the X-ray band with persistent luminosities, $L_{\rm
X} \approx 10^{34}-10^{36}$ erg s$^{-1}$, which are orders of magnitude higher
than their rotation power for most sources. Soft X-ray spectra of magnetars
below 10 keV are soft and can be empirically described by either the sum of two
blackbody (BB) functions, or by the sum of a BB ($kT \sim 0.5$ keV) and  a
power-law (PL) model with a photon index of $\Gamma \sim 2-4$
\citep{mereghetti15}. In addition to their soft X-ray emission, some magnetars
also persistently emit bright hard X-rays, detected with {\it INTEGRAL}, {\it
RXTE}, \citep{kuiper06, den08a, den08b}, {\it SUZAKU}, \citep{enoto10}, and
recently with the {\it Nuclear Spectroscopic Telescope Array} \citep[{\it
NuSTAR}; see, e.g.,][]{vogel14,an15}. The hard X-ray spectra can be fit by a PL
with typical photon indices of $\Gamma \sim 1-1.5$ (see \citealt{rea11}, and
\citealt{mereghetti15} for recent reviews).

Unlike magnetospheres in ordinary radio pulsars, those of magnetars are
expected to be highly twisted due to the deformation of the NS crust
\citep{thompson02}. The twisted magnetospheres can support dense charge
particles which further provide large optical depth ($\sim 1-10$) to resonant
cyclotron scattering (RCS). In the magnetar framework, soft X-ray spectra are
generally explained as the NS surface thermal emission Compton up-scattered by
the plasma in the magnetosphere which form a PL tail \citep[see
e.g.,][]{lyutikov06, fernandez07, tong10}. Both one-dimensional (1D) and
three-dimensional (3D) RCS models had been proposed \citep[see,
e.g.,][]{lyutikov06, fernandez07, nobili08a}, and applied to the observed
spectra \citep{rea08, zane09}. However, the scattering process is insensitive
to the surface magnetic field; therefore, one cannot deduce magnetospheric
properties via modeling with the RCS process.

In addition to the RCS in twisted magnetospheres, the vacuum polarization and
proton cyclotron resonances play key roles in the propagation of radiation
through magnetar atmospheres, leaving signatures of surface magnetic fields in
the continuum spectra of magnetars \citep{ozel01,ozel03,ho03}. Taking into
account the combined effects of the magnetar atmosphere and reprocessing in the
magnetosphere, \cite{guver07} developed the 1D surface thermal emission and
magnetospheric scattering (STEMS) model, which offers acceptable fits to the
soft X-ray spectra of numerous magnetars \citep{guver07, guver08, ng11, lin12}.
Recently, we considered the same physical processes but calculated the
synthetic spectra using a Monte Carlo technique (Weng et al. 2015, hereafter
Paper I). The set of calculated spectra are further used to create a tabular
model (STEMS3D) that can be accommodated by the X-ray fitting package XSPEC.
Investigating a large sample of {\it XMM-Newton} spectra of magnetars, we found
that the STEMS3D model yielded successful fits to nearly all of the spectra
with a reduced $\chi^{2}$ less than 1.3. We concluded that the STEMS3D model
provides a robust method to measure the surface magnetic field strength and the
magnetospheric configuration, that is, the level of magnetospheric twist. The
results of modeling the soft X-ray spectra of magnetars with STEMS3D showed
that the magnetosphere in a magnetar is highly twisted, and at least in one
case (\de2259 in its 2002 outburst), we most likely observe an indication of
the untwisting of a twisted magnetosphere.

At present, hard X-ray emission has been reported in about one-third of
magnetars. There is still no clear evidence of the peak energy beyond which the
hard X-ray emission would decline rapidly \citep{den08a, den08b}. Nevertheless,
there was an indirect indication of $\sim 1-2$ MeV for the spectral break
energy \citep{sasmaz10}. Additionally, hard X-ray emission in transient sources
exhibits time variations, as in the 2009 outburst of 1E 1547.0-5408
\citep{kuiper12}. \cite{beloborodov13} attributed the hard X-ray emission to
the interaction of surface emission with the relativistic particles surrounding
magnetars, which also could make a non-negligible contribution to the emission
below 10 keV. The observed non-relativistic nature of magnetospheric particles
(electron velocity, $\beta \leq 0.3$) as presented in Paper I places a
constraint on the viability of such a mechanism. However, it is important to
note the fact that electrons with mildly relativistic velocities, $\beta \geq
0.6$, could effectively scatter soft photons up to $\sim 100$ keV. It is
therefore crucial to diagnose the scattering scenario using a broadband
spectral modeling.

{\it NuSTAR}, the first focusing hard X-ray instrument \citep{harrison13},
executed a series observations of a small set of bright magnetars. Here, we
employed  {\it NuSTAR} observations of four magnetars: \axpu0142, \ce1841,
\de2259, and \ae1048, together with simultaneous or contemporaneous soft X-ray
observations to perform broadband spectral analysis using the expanded STEMS3D
model. In the next section, we briefly describe the STEMS3D model and the
Monte-Carlo-based methodology used to extend its spectral coverage into hard
X-rays. We present the results of broadband spectral investigations using the
expanded STEMS3D in Section 3. We discuss the implications of these results and
present our main conclusions in Section 4.

\section{Three-dimensional surface thermal emission and magnetospheric scattering model}
In this section, we provide a brief overview of the STEMS3D model as further
details of the model were already presented in Paper I. The STEMS3D model
envisions that the emergent radiation of magnetars is significantly affected by
their atmospheres and twisted magnetospheres. The emission from the NS surface
is expected to become distorted while propagating through the ionized, highly
magnetized atmospheres by the absorption, emission, and scattering processes
\citep{ozel03}. In turn, proton cyclotron resonance gives rise to a line
feature while the vacuum polarization weakens this line feature and also
enhances the conversion between photons of different polarization modes
\citep{ozel03}. The spectral profile of the emerging surface emission is
described by the surface temperature ($kT$) and magnetic field \citep[$B$; see
Figure 4 in][]{ozel03}. These emerging photons would further interact with the
charged particles in the magnetosphere by multiple resonant scattering,
manifesting themselves as the high-energy tail \citep{guver07, nobili08a,
rea08}.

The STEMS3D model has four model parameters: surface temperature $kT$, magnetic
field strength at the pole $B$, twist angle $\Delta\phi$, and the electron
velocity $\beta$. Unlike the previous version of STEMS3D presented in Paper I,
we now restricted the model parameter space of $kT \leq 0.5$ keV, and
$\Delta\phi \geq 0.5$ radian, as suggested by the soft X-ray spectral fitting
(see Paper I). In these cases, a significant fraction of the seed photons are
scattered up to the hard X-ray band with moderate optical depths ($\gtrsim
0.5$, Figure 3 in Paper I), which is a few orders larger than the tiny fraction
contributed by the surface emission above 10--15 keV. Therefore, seed photons
at energies above this level would not influence the emerging spectrum, hence,
they can safely be discarded. We calculated more than 7000 model spectra in the
0.1-100 keV range, above which the electron recoil effect becomes important
\citep{nobili08b}. The parameter ranges for the current version of STEMS3D are
$kT = 0.1-0.5$ keV (with a step of 0.1 keV), $B = 10^{14}-10^{15}$ G (with a
step $10^{14}$ G), $\Delta\phi = 0.5-2.0$ radian (with a step of 0.1 radians),
and $\beta$ = 0.1-0.9 (with a step of 0.1).

\begin{figure*}
\begin{center}
\includegraphics[scale=0.4]{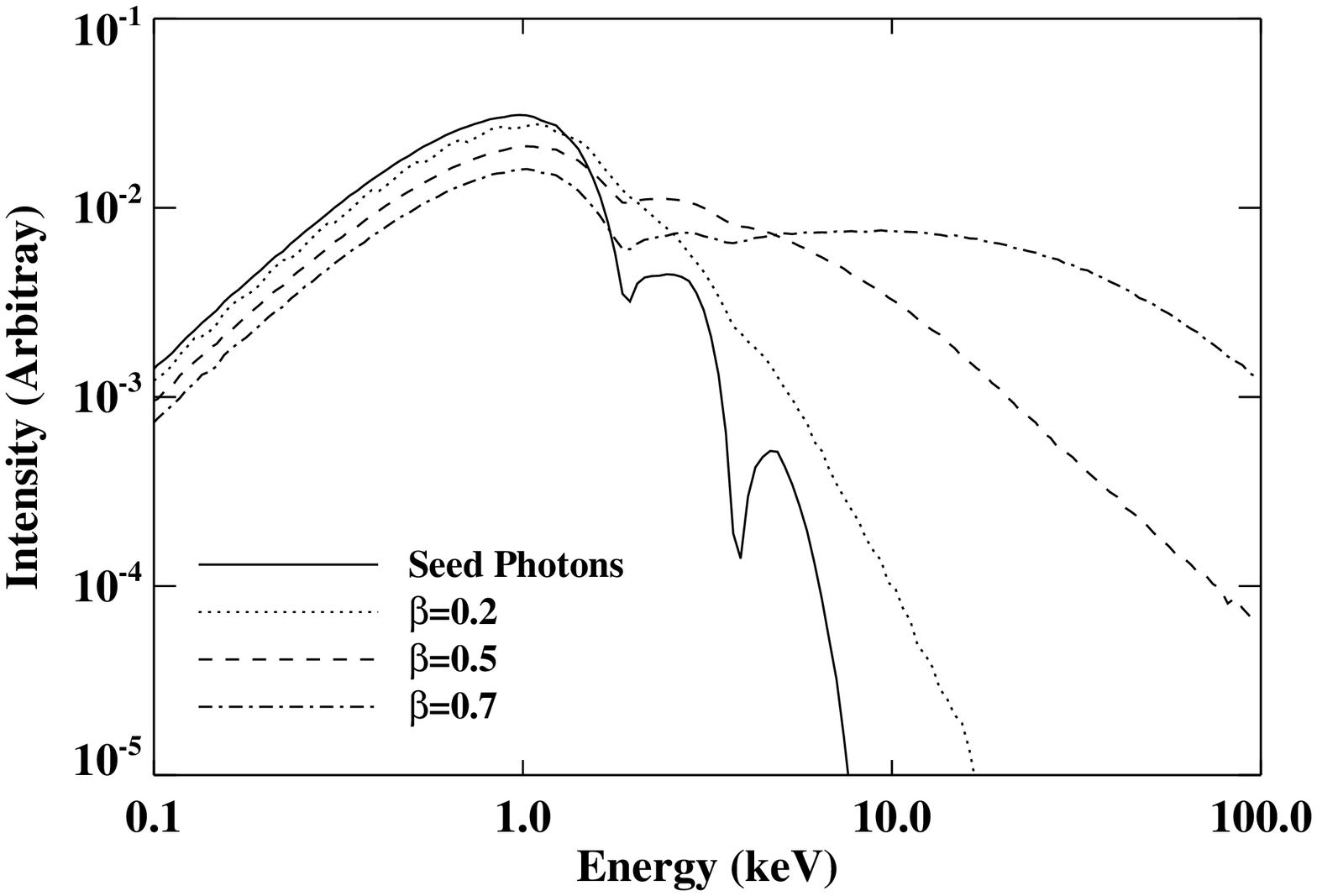}
\includegraphics[scale=0.4]{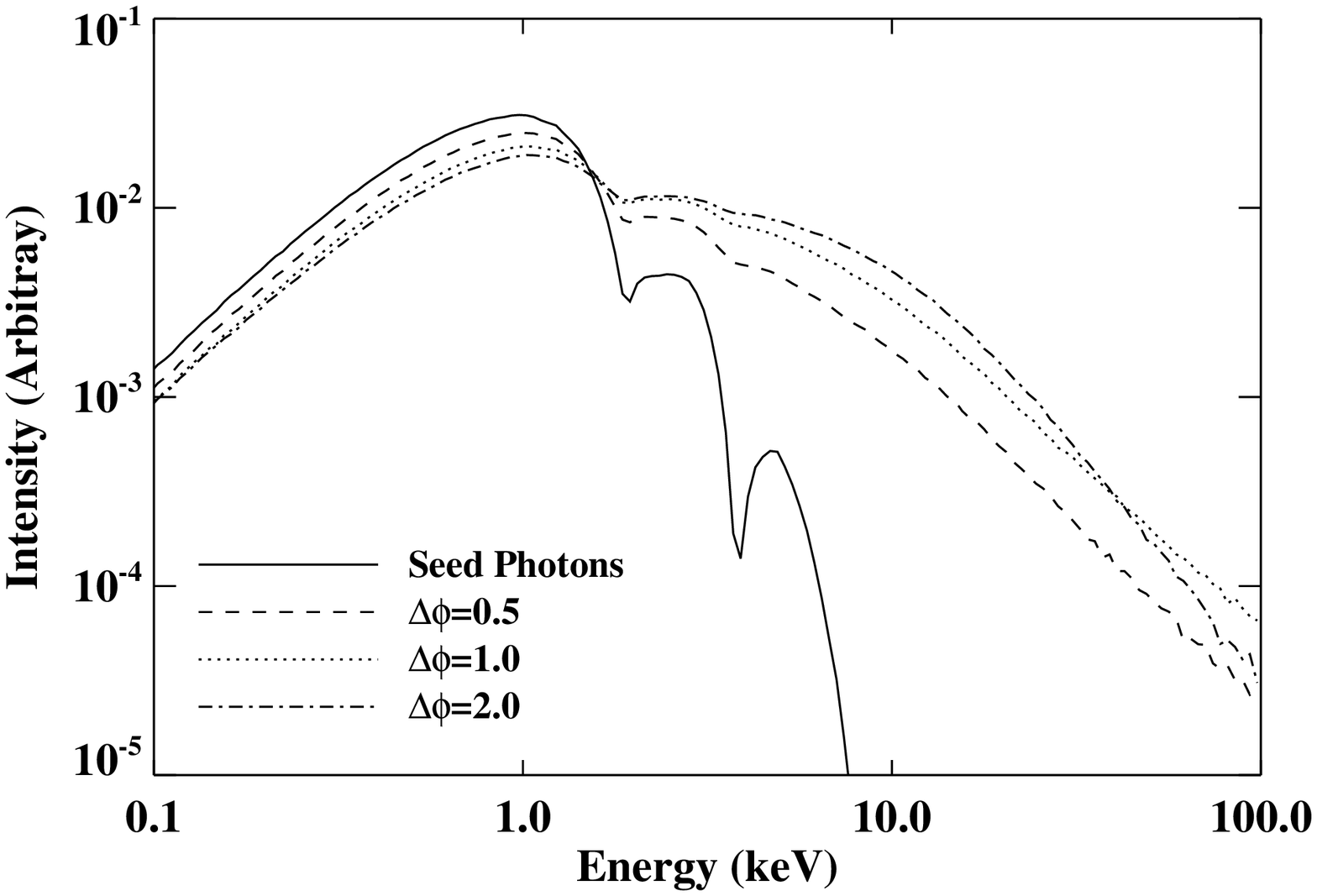}\caption{Left: model spectra for
surface magnetic field strength, $B = 3\times10^{14}$ G, surface temperature,
$kT = 0.3$ keV, magnetospheric twist angle, $\Delta\phi = 1.0$, and three
different values of magnetospheric electron velocity, $\beta$. Right: STEMS3D
spectra for $B = 3\times10^{14}$ G, $kT = 0.3$ keV, $\beta = 0.5$ and three
different values of $\Delta\phi$. \label{spectra}}
\end{center}
\end{figure*}

In the STEMS3D model, the current density and the optical depth are
self-consistently derived from the $\Delta\phi$ and the $\beta$. Consequently,
the Compton up-scattering efficiency, and therefore, the spectral profile, is
sensitive to these two parameters. We plot the STEMS3D spectra in Figure
\ref{spectra} to illustrate the effects of $\Delta\phi$ and the $\beta$. As can
be seen, the proton cyclotron lines are weakened by multiple scatterings in the
twisted magnetosphere, and the hard X-ray flux level increases with
$\Delta\phi$ as well as $\beta$. When a magnetosphere becomes optically thick,
a larger fraction of seed photons is scattered to the energy around the
electron energy, forming a second hump. Thus, the photon density beyond the
electron energy looks suppressed in the normalized spectra with a larger
optical depth ($\Delta\phi \sim 2$, see also Figure 5 in Nobili et al. 2008a).
Such a trend in evolution is shown in other RCS models as well, e.g., 1D RCS
\citep[Figure 1 in][]{rea08}.

\section{Observations, Data Analysis \& Results}

{\it NuSTAR} consists of two focusing instruments and two focal plane modules,
called Focal Plane Modules A and B (hereafter FPMA and FPMB). The instruments
are sensitive to photon energies in the range of 3--79 keV. Soft X-ray
telescopes (3--10 keV) are useful for cross calibration. {\it NuSTAR} is
currently the most sensitive detector in the 10--79 keV range due to its unique
focusing ability; thus, it can acquire high-quality data from magnetars within
a relatively short exposure time. To date, {\it NuSTAR} has observed five
magnetars: \axpu0142 \citep{tendulkar15}, \ce1841 \citep{an15}, \de2259
\citep{vogel14}, \ae1048 \citep{an14}, and the Galactic Center magnetar SGR
1745--2900 \citep{mori13, kaspi14}. There have also been simultaneous or
contemporaneous pointed X-ray observations of these sources with {\it
Swift}/XRT or {\it XMM-Newton}. In our investigations, we used the {\it NuSTAR}
data of four magnetars and contemporaneous X-ray observations for broadband
spectral analysis (see Table \ref{log}). We excluded the Galactic Center
magnetar from our sample due to the very high interstellar absorption in its
galactic vicinity.

We processed {\it NuSTAR} and {\it Swift} with HEAsoft version 6.16. In
particular, {\it NuSTAR}  \citep{harrison13} data were processed with
\texttt{nupipeline} and CALDB version 20141107 using standard quality
thresholds. The source spectra were extracted from a circular region ($R =
60\arcsec$) with the task \texttt{nuproducts} and the resulting spectra were
grouped to have at least 50 counts per spectral bin. In each pointing, we chose
a source-free region to extract the background spectrum, except for \ce1841
(see below). For {\it Swift} XRT \citep{gehrels04} data, the initial event
cleaning was performed with the task \texttt{xrtpipeline} using standard
filters. To ensure reliable results using $\chi^{2}$ statistical analysis, the
spectra were grouped to have at least 30 counts per spectral bin. For {\it
XMM-Newton} observations, the data collected with EPIC instruments
\citep{struder01} were reduced using the Science Analysis System (SAS) software
version 12.0.1, and filtered with standard criteria. We provide further source
specific processing details below. All of the spectra were then fitted with
XSPEC 12.8.2 \citep{arnaud96}. We assume a gravitational redshift correctionof
0.306 for the emergent emission, corresponding to an NS with mass 1.4
$M_{\odot}$ and $R_{\rm NS} = 10$ km, and we adopt the solar abundances by
\cite{lodders03}.

\begin{deluxetable*}{clccccc}
\tabletypesize{\tiny} \tablewidth{0pt} \tablecaption{Log of observations}
\tablehead{\colhead{Source}& \colhead{Observatory/} & \colhead{ObsID} &
\colhead{Obs Date} & \colhead{Mode} & \colhead{Net Exposure} &
\colhead{Energy Band} \\
\colhead{} & \colhead{Instrument} & \colhead{} & \colhead{} &\colhead{} &
\colhead{(ks)} & \colhead{(keV)}} \startdata \hline \\
\axpu0142     & {\it Swift}  & 00080026001 & 2014 Mar 27 & WT & 5 & 1-10 \\
     ...      & {\it Swift}  & 00080026002 & 2014 Mar 28 & WT & 13 & 1-10 \\
     ...      & {\it Swift}  & 00080026003 & 2014 Mar 28 & WT & 7 & 1-10 \\
     ...      & {\it NuSTAR} &  30001023002 & 2014 Mar 27 & Science & 24 & 3-79 \\
     ...      & {\it NuSTAR} &  30001023003 & 2014 Mar 28 & Science & 144 & 3-79 \\
\hline \\
\ce1841       & pn  &  0013340101 & 2002 Oct 05 & LW & 2 & 1-10 \\
     ...      & pn  &  0013340201 & 2002 Oct 07 & LW & 4 & 1-10 \\
     ...      & {\it NuSTAR} &  30001025012 & 2013 Sep 21 & Science & 101 & 3-79 \\
\hline \\
\de2259       & {\it Swift}& 00080292002 & 2013 Apr 25 & PC  & 13 & 0.5-10 \\
     ...      & {\it Swift}& 00080292003 & 2013 Apr 26 & PC  & 14  & 0.5-10\\
     ...      & {\it Swift}& 00080292004 & 2013 Apr 28 & PC  & 14  & 0.5-10\\
     ...      & {\it NuSTAR} &  30001026002 & 2013 Apr 24 & Science & 37 & 4-79 \\
     ...      & {\it NuSTAR} &  30001026003 & 2013 Apr 25 & Science & 15 & 4-79 \\
     ...      & {\it NuSTAR} &  30001026005 & 2013 Apr 26 & Science & 16 & 4-79 \\
     ...      & {\it NuSTAR} &  30001026007 & 2013 May 16 & Science & 88 & 4-79 \\
\hline \\
\ae1048       & pn/MOS1/MOS2 &   0723330101 & 2013 Jul 22 & FF/SW/SW & 53/72/71 & 0.5-10\\
     ...      & {\it NuSTAR} &  30001024002 & 2013 Jul 17 & Science & 25  & 3-12  \\
     ...      & {\it NuSTAR} &  30001024003 & 2013 Jul 17 & Science & 22  & 3-12  \\
     ...      & {\it NuSTAR} &  30001024005 & 2013 Jul 19 & Science & 153 & 3-12  \\
     ...      & {\it NuSTAR} &  30001024007 & 2013 Jul 25 & Science & 104 & 3-12
\enddata

\tablecomments{Net exposure: clean exposure after background flares and short
bursts excluded. Mode: operating mode including Windowed Timing (WT),
Photon Counting (PC), Full Frame Window (FF), Large Window (LW), and Small
Window (SW). \label{log}}
\end{deluxetable*}

\begin{figure*}
\begin{center}
\includegraphics[scale=0.5]{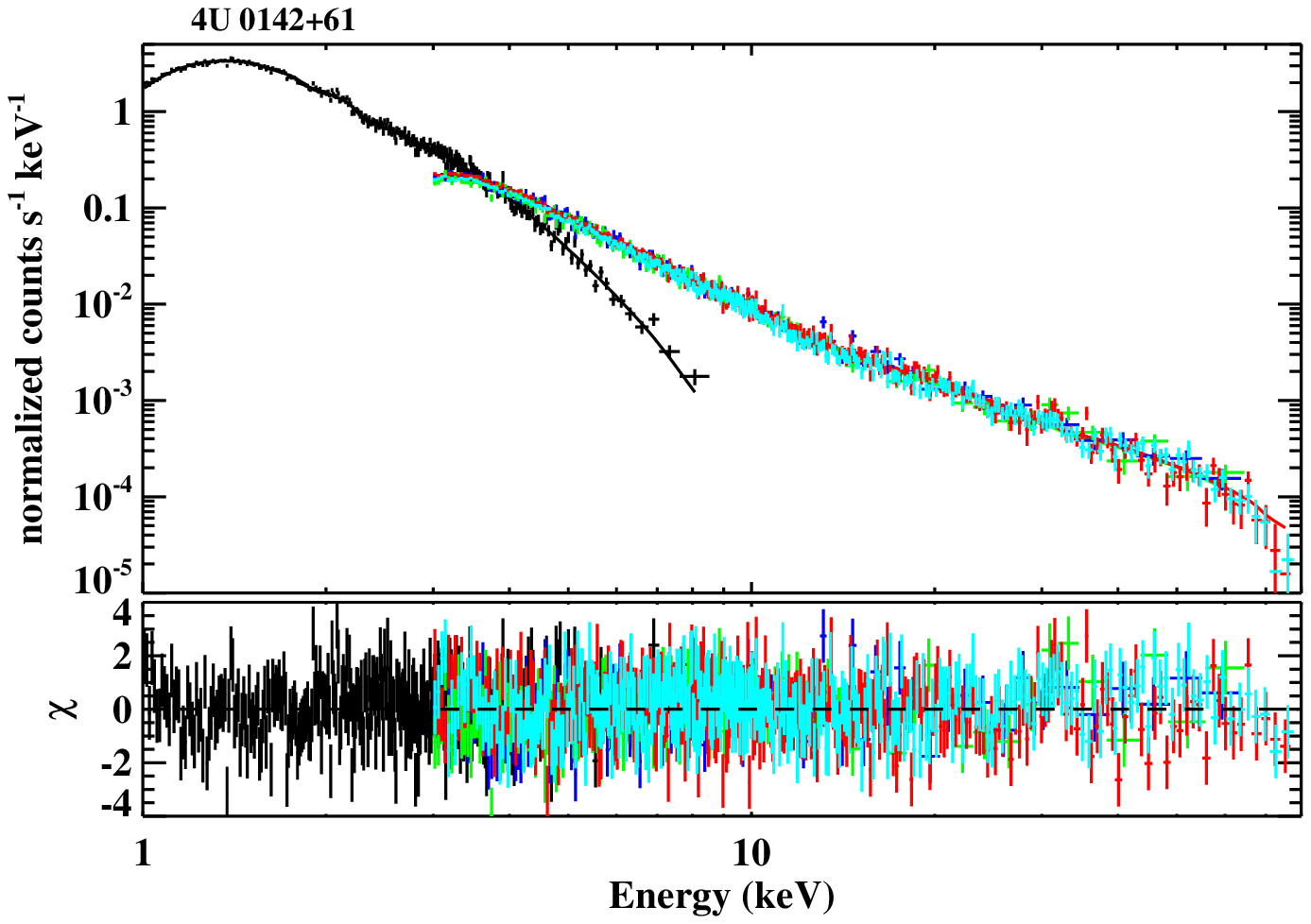}
\includegraphics[scale=0.5]{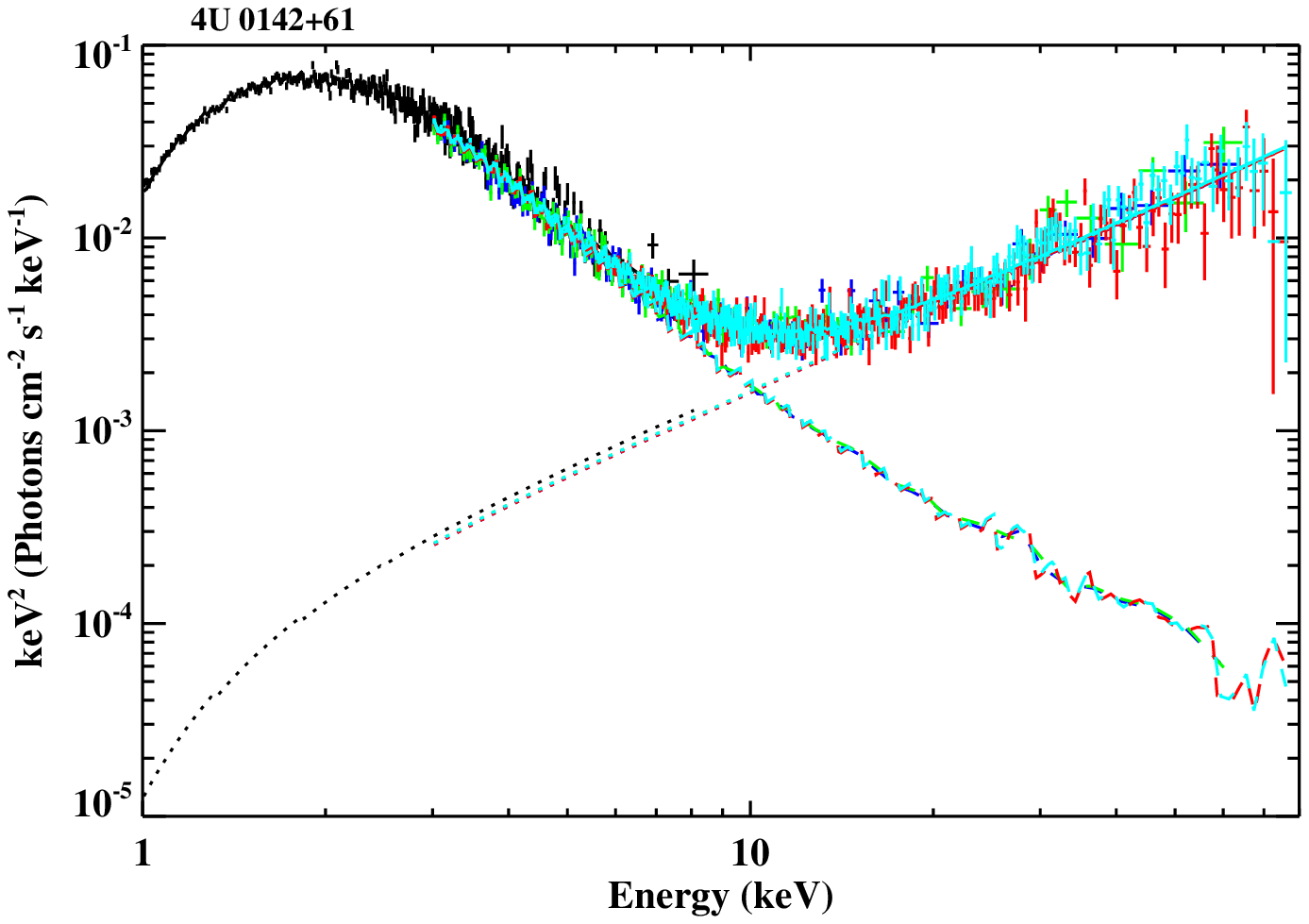}
\includegraphics[scale=0.5]{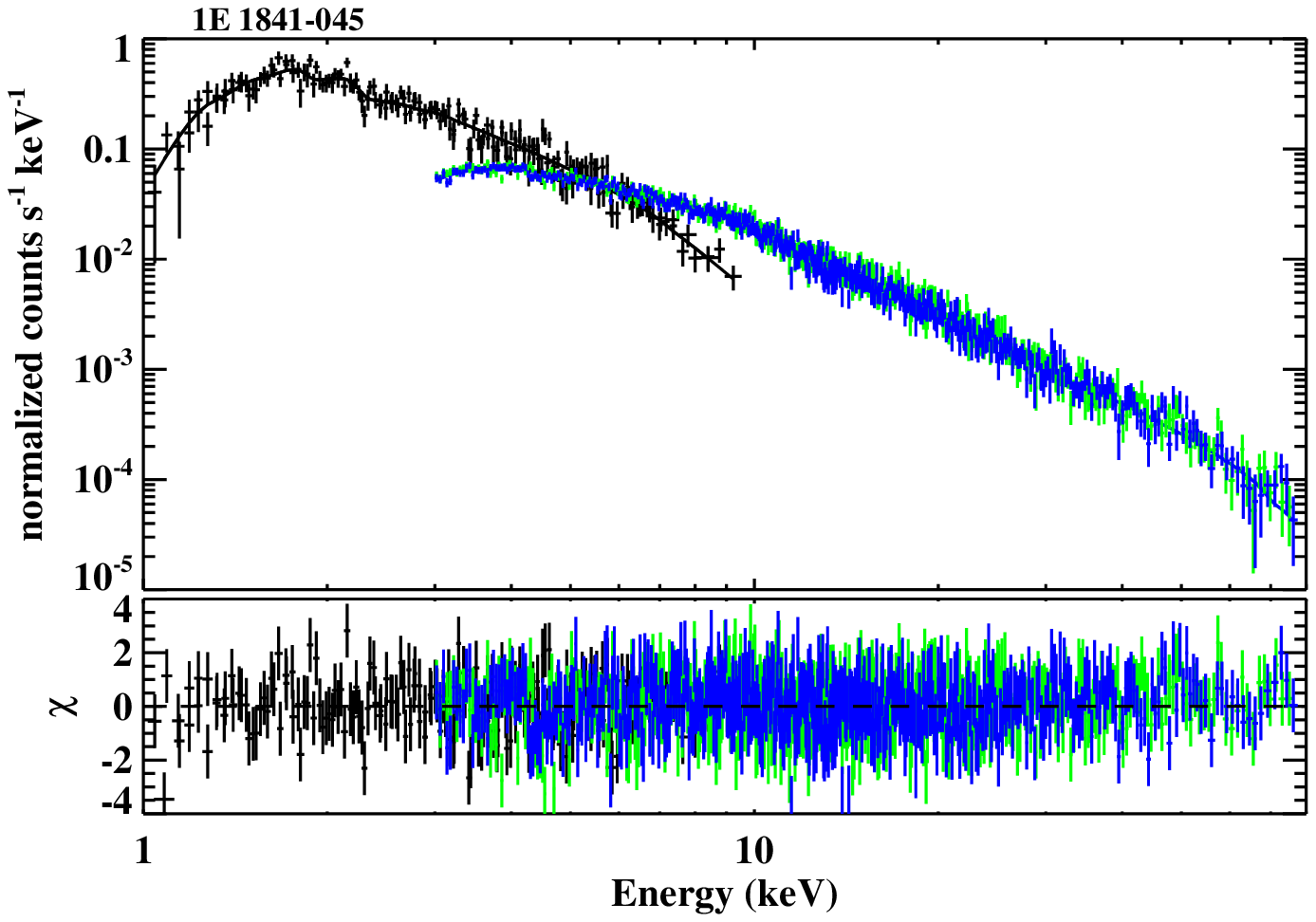}
\includegraphics[scale=0.5]{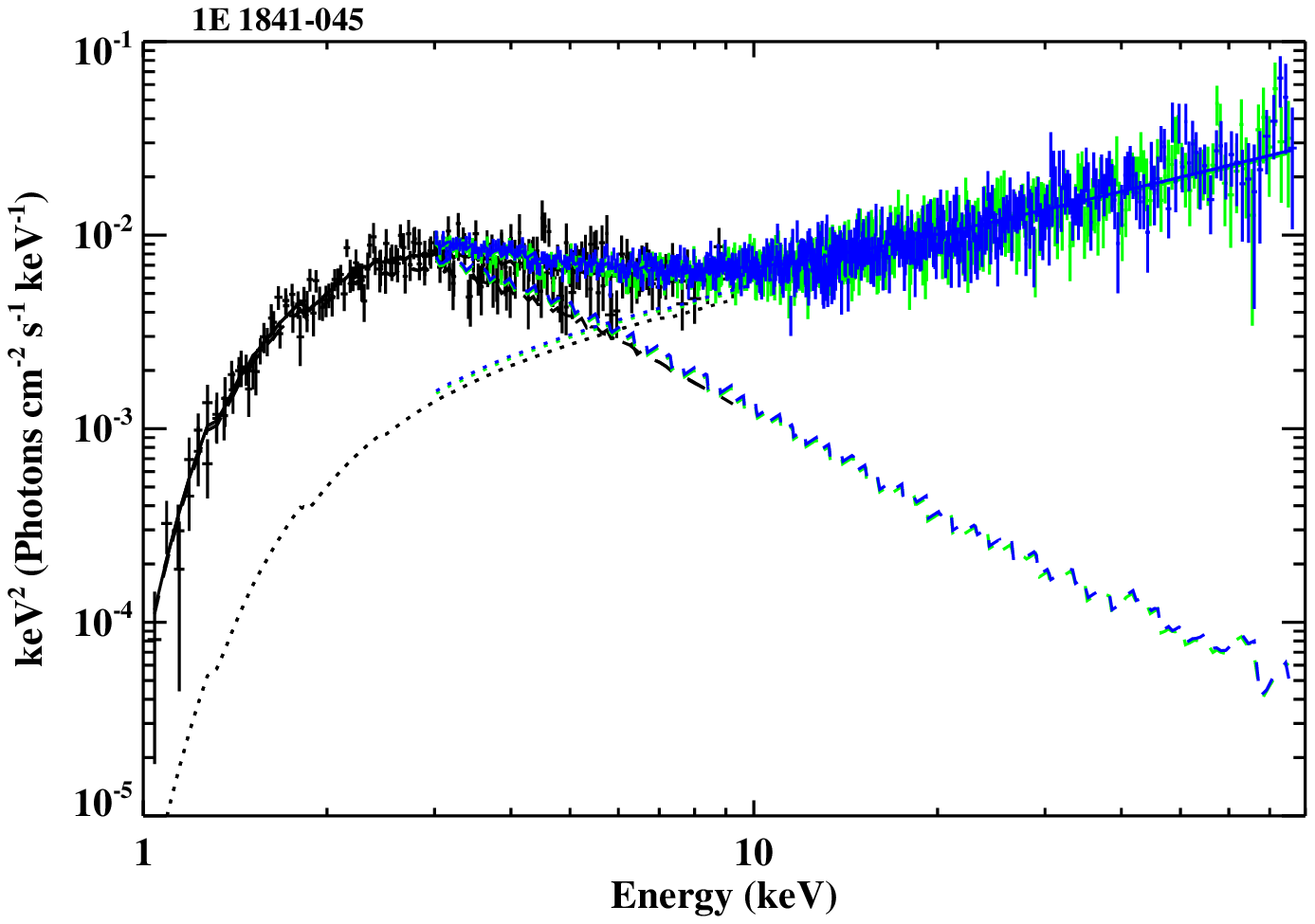}
\includegraphics[scale=0.5]{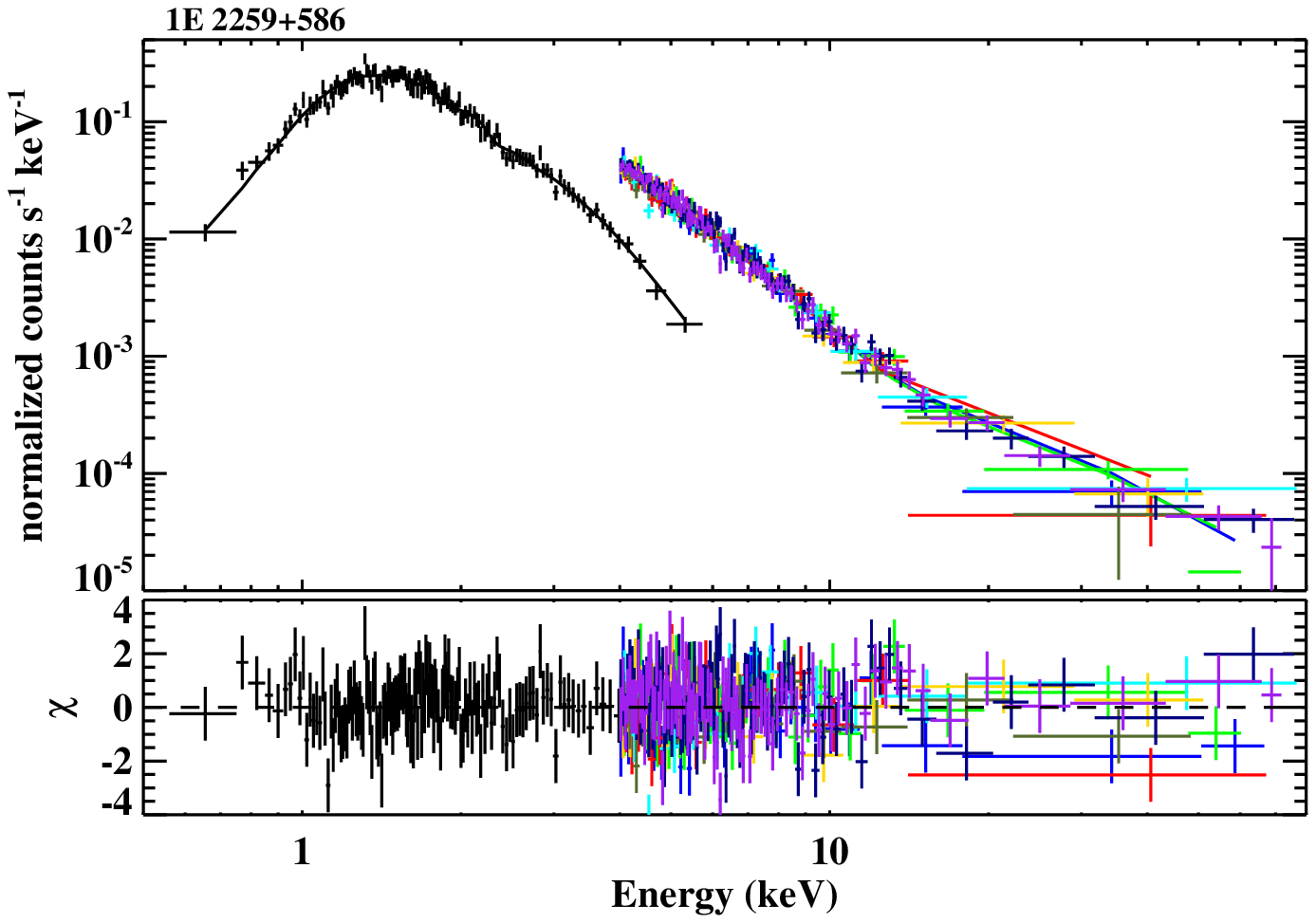}
\includegraphics[scale=0.5]{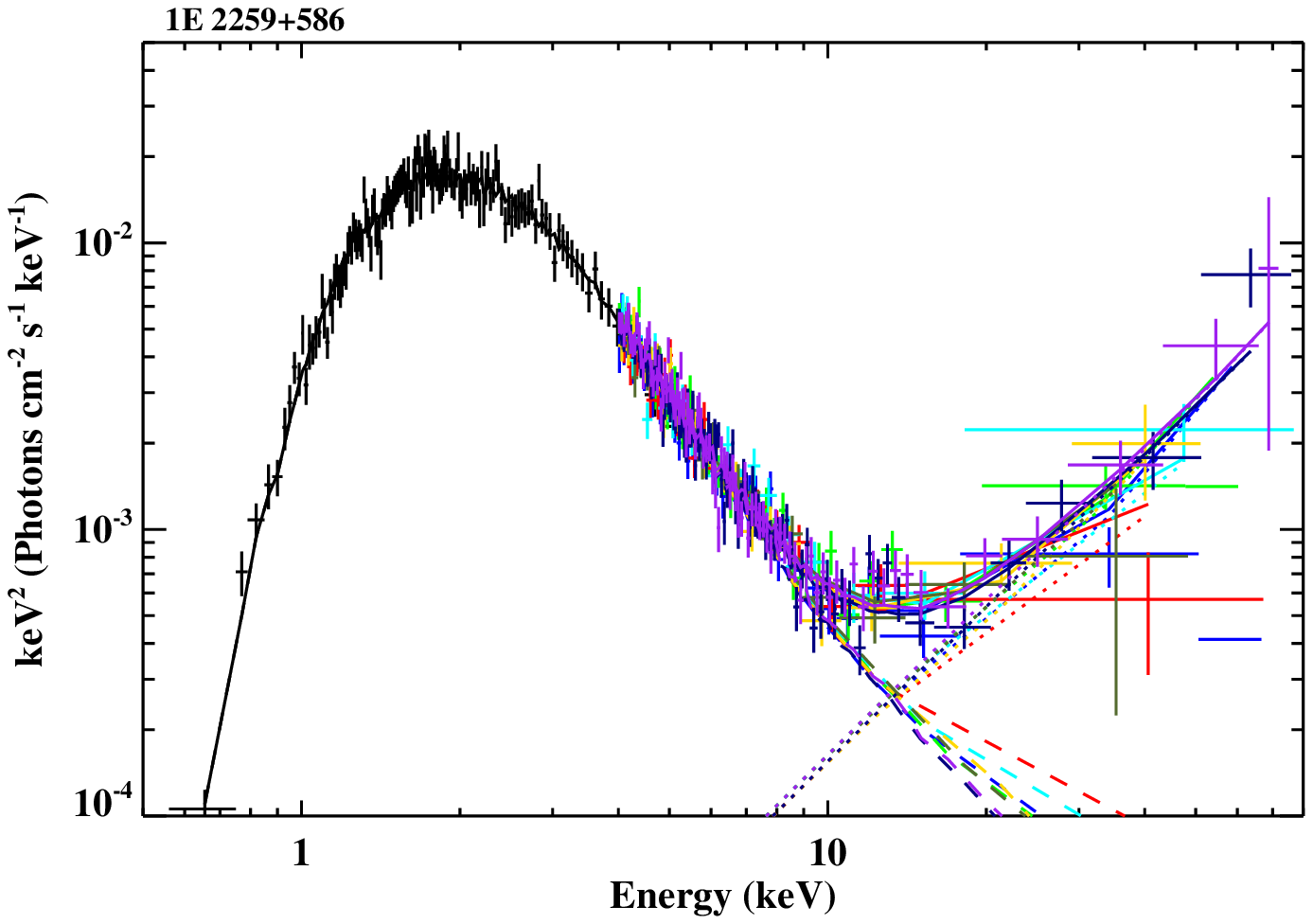}
\includegraphics[scale=0.5]{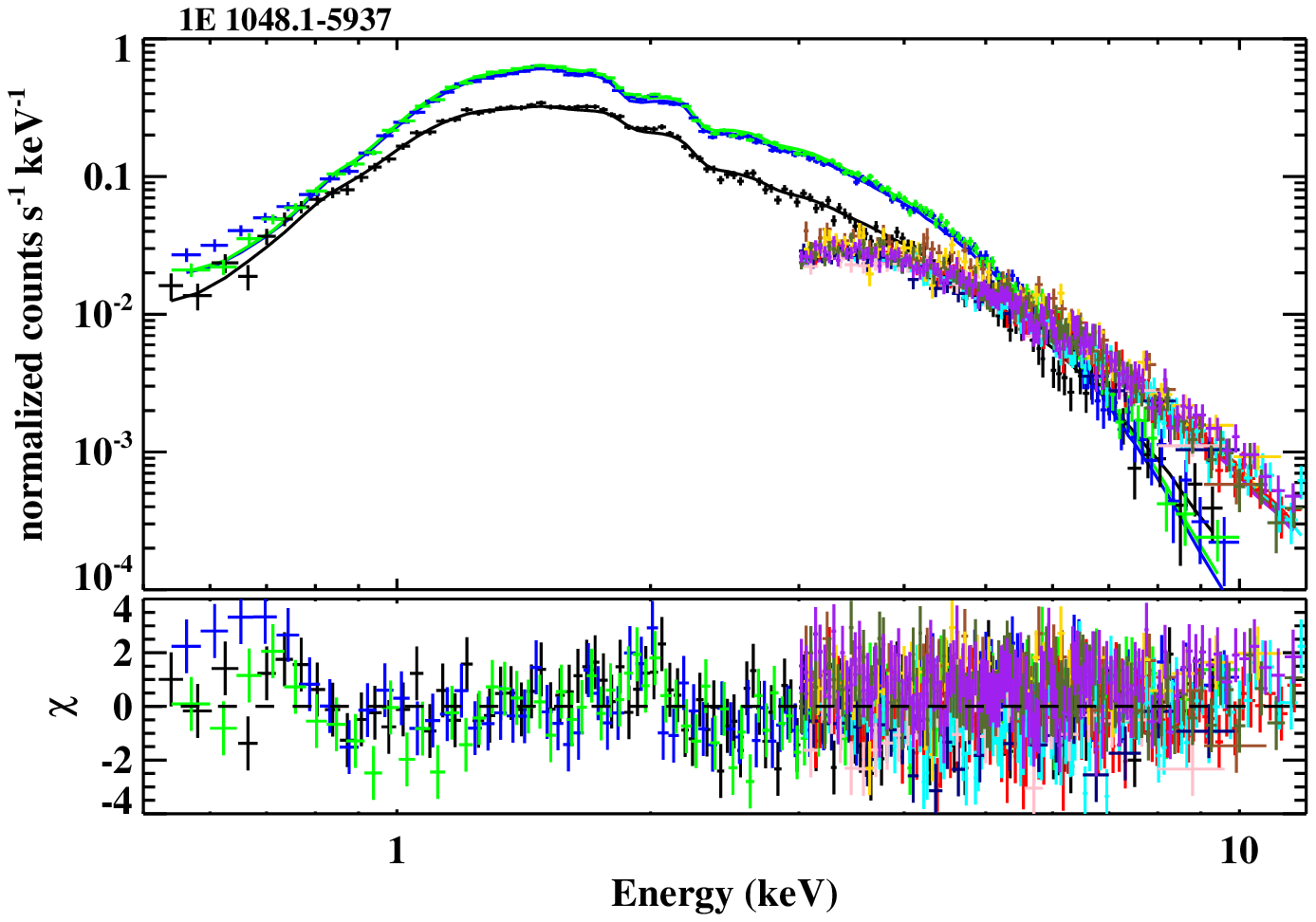}
\includegraphics[scale=0.5]{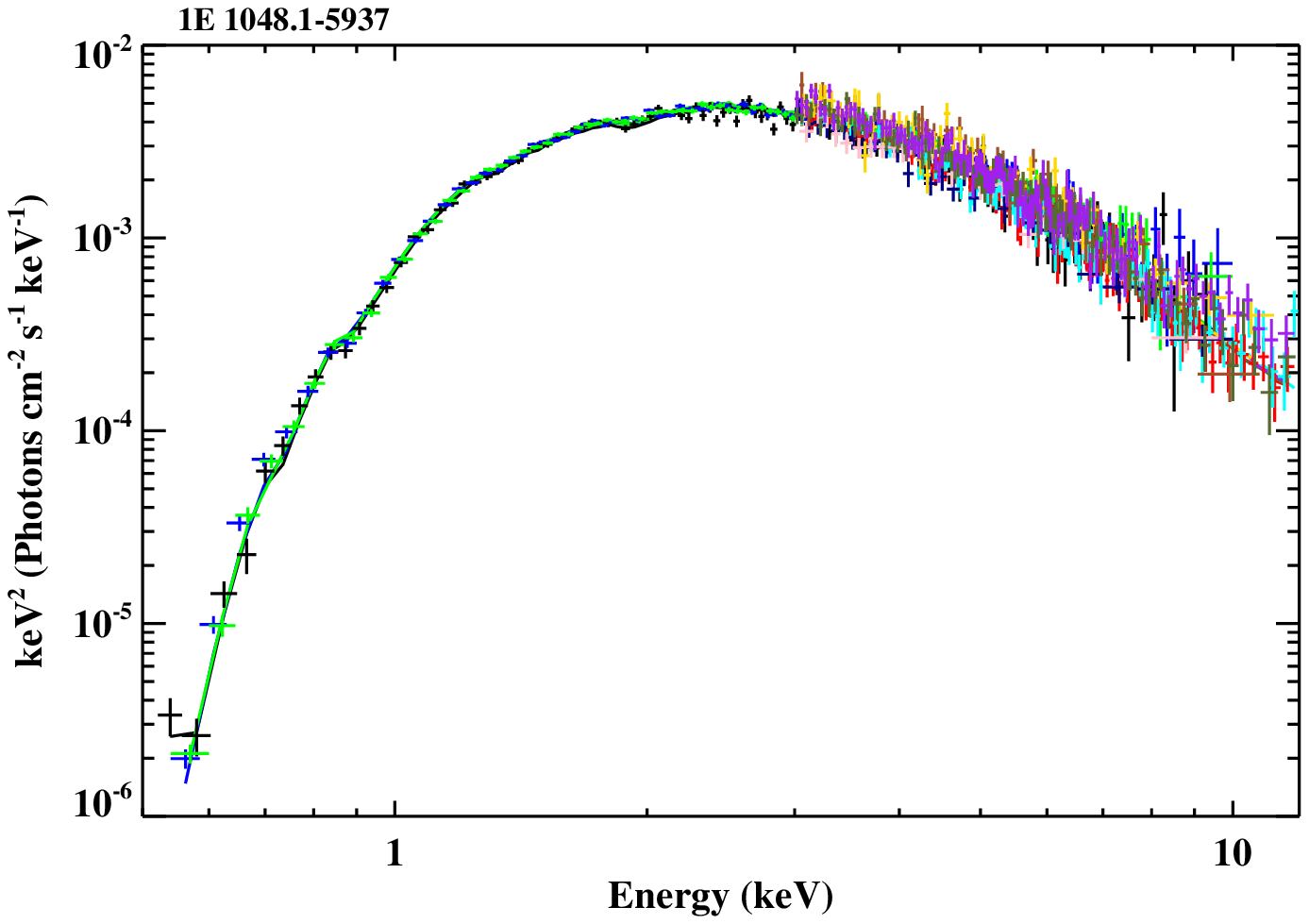}

\caption{Left panels: X-ray count spectra of \axpu0142, \ce1841, \de2259 and
\ae1048 and the best-fitting STEMS3D model curves. The fit residuals are
displayed below each spectrum. Right panels: the corresponding photon spectra
of the four sources. The dotted, dashed, and solid lines in the right panels
represent the STEMS3D, PL, and the sum of the two components,
respectively.\label{spec}}
\end{center}
\end{figure*}

\axpu0142 is the brightest AXP and its X-ray emission is quite steady in both
the hard and soft X-ray bands. There were two {\it NuSTAR} and three {\it
Swift} observations executed at the end of 2014 March. {\it Swift} XRT
observations were performed in Windowed Timing (WT) mode to avoid pile up. We
extracted both the source and background signals from 50-pixel long strips.
Since there were the model-independent residuals were exhibited below 1 keV, we
fit the {\it Swift} spectrum in the range of 1--10 keV together with {\it
NuSTAR} spectra in the 3-79 keV range. In this paper, we fit all of the
available spectra for each source simultaneously by linking model parameters
together so that they can converge to common values while allowing model
normalizations to vary. We also included a multiplicative constant for each
model component to account for cross-normalization issues, which were fixed to
unity for data collected with the soft X-ray missions, and allowed to vary for
FPMA and FPMB. In all of the cases, we find that the cross-normalization
factors are not far from unity, ranging between 0.88 and 1.12. We find that the
broadband X-ray spectrum of \axpu0142 is very well described by the STEMS3D +
PL model ($\chi^{2}$/dof = 1290.3/1235), yielding a magnetospheric twist angle
of 1.7 rad and magnetospheric electron velocity of 0.21 $c$, which are in
agreement with what we obtained using {\it XMM-Newton} observations only (Paper
I). In the top row of Figure \ref{spec}, we present combined \axpu0142 data
sets together with best-fit model curves and fit residuals, as well as spectral
decomposition of each model component. The hard X-ray component represented by
a PL has an index of \textbf{$0.57_{-0.05}^{+0.04}$} which is flatter than what
is observed with {\it INTEGRAL} (\citealt{den08a}, but see \citealt{wang14}).

\ce1841, located at the center of supernova remnant (SNR) Kes 73
\citep{vasisht97}, is another steady X-ray emitter. The source was monitored
with {\it NuSTAR} between 2013 September 5 and 21, during which time six bursts
were detected \citep{an15}. In order to eliminate the contamination from the
SNR, an annulus region with inner and outer radii of 60\arcsec and 100\arcsec
around the source position was selected for background spectral extraction, as
was also done in \citet{an13}. We searched for soft X-ray data from 2013 August
1 to 2013 October 30, and identified 11 {\it Swift} XRT observations, 10
executed in WT mode and 1 short observation in Photon Counting (PC) mode. Note
that in WT mode, 10 rows are compressed into a single row, only the central 200
columns are read, and only 1D imaging is preserved. consequently, it is
impossible to resolve the SNR contribution and three SNR related emission lines
(Mg, Si, and S) clearly appear in the stacked spectrum of 10 WT data. Since the
source emission has been stable for more than 15 years \citep{dib14}, we
performed the joint fit with the longest, and burst-free {\it NuSTAR}
observations (ObsID = 30001025012) and {\it XMM-Newton} observations that were
performed $\sim 11$ years earlier. For the {\it XMM-Newton} observation, since
the MOS1/MOS2 data were performed in the full frame mode and were seriously
affected by pile up, we only use the pn data in the following spectral fitting.
The joint spectra are fit by our model (STEMS3D+PL) and a 2\% systematic error
is added to the {\it XMM-Newton} data to account for calibration uncertainties.
The fit results yielded that \ce1841 has the lowest twist angle ($\Delta\phi
\sim 0.83$) and the highest electron velocities in the magnetosphere ($\beta
\sim 0.28$) among the four sources studied here. The resulting photon index of
the PL component ($\Gamma \sim 1.23$) is consistent with the value reported
based on {\it INTEGRAL} data \citep{kuiper06}. However, due to the short
exposure of {\it XMM-Newton} data, in addition to the high interstellar
absorption, the parameter values have large uncertainties (see Table \ref{fits}
and Figure \ref{spec}). Note that \ce1841 is the only source whose non-thermal
component ($F_{\rm P} \sim 5.1\times 10^{-11}$ erg s$^{-1}$ cm$^{-2}$)
dominates the STEMS3D component ($F_{\rm S} \sim 3.6\times 10^{-11}$ erg
s$^{-1}$ cm$^{-2}$) in the range of 1--79 keV.

\de2259 is considered to be the AXP prototype and its soft X-ray spectrum is
typical of AXPs, empirically described by a BB and a PL components
\citep{gavriil04}. Fitting the {\it XMM-Newton} spectra collected during its
2002 outburst with the STEMS3D model, we found that an increased twist angle
was coincident with the outburst and then decreased subsequently during the
outburst decay, which agrees perfectly with the twisting/untwisting
magnetosphere scenario (Paper I). Persistent emission above 10 keV had been
detected in \de2259 with {\it RXTE} \citep{kuiper06}, and pulsations above 20
keV were first revealed in {\it NuSTAR} data \citep{vogel14}. Here, we analyzed
all four {\it NuSTAR} observations. \de2259 is located near the center of the
SNR CTB 109, which has a half-shell morphology \citep{gregory80}. Investigating
a {\it Chandra} data, \cite{sasaki04} pointed out that no emission above 4 keV
was emitted by CTB 109. Thus, we only use the {\it NuSTAR} spectra above 4 keV
to minimize the contamination from the SNR. {\it Swift} XRT observations were
performed in PC mode, and its angular resolution is good enough to resolve the
SNR. Because of the fact that XRT data were suffered from the pile-up effect,
we extracted source spectra within an annular region centered on the source
position (R.A. = 23:01:08.1, decl. = +58:52:44.5, J2000)  with the inner radius
of 5 pixels and create ARF files to correct for the loss of counts caused by
annular exclusion. The fitted parameters of the STEMS3D model are B $\sim
5.6\times10^{14}$ G, $\Delta\phi \sim 1.6$, and $\beta \sim 0.2$. In contract
to \ce1841, the broadband {\it NuSTAR} spectra of \de2259 yield a very hard
($\Gamma \sim 0.2$) but weak hard X-ray component, contributing a fraction of
$\sim 10\%$ of total X-ray emission.

\begin{figure}
\begin{center}
\includegraphics[scale=0.5]{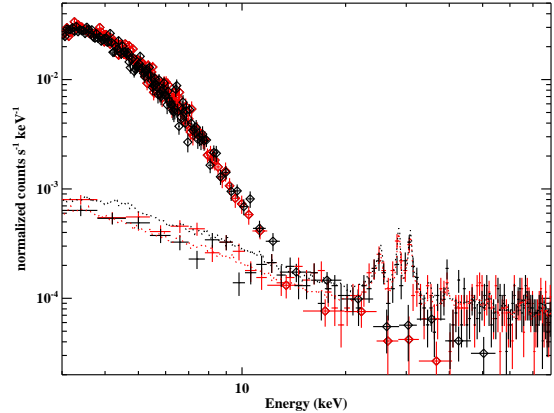}\caption{{\it NuSTAR} FPMA/FPMB
spectra (diamonds, ObsID=30001024005) of \ae1048 and their corresponding
background spectra (crosses). We also plot the background spectra generated
with \texttt{nuskybgd} but without error bars for clarity (dotted lines). Data
from FPMA and FPMB are marked with red and black symbols, respectively.
\label{bkg}}
\end{center}
\end{figure}

The AXP \ae1048 displays bright and variable soft X-ray emission, but its hard
X-ray radiation was only observed during short bursts. Recently, \cite{an14}
reported the detection of eight bursts during the 2013 July observation
campaign with the {\it NuSTAR}. After removing the times of the burst
intervals, we do not detect any significant emission above 15 keV (Figure
\ref{bkg}) and the persistent emission below 15 keV remains constant during
different observations. For the simultaneous {\it XMM-Newton} observation
(ObsID=0723330101, on July 22), the pn data were executed in the full-frame
mode and suffered from the pile up, while the MOS1 and MOS2 data were taken in
the small window mode and thus avoided the pile up. After correcting the
pile-up effect for the pn data, we performed simultaneous fit with the {\it
NuSTAR} observations and {\it XMM-Newton} observation performed in 2013 July.
Fitting the joint spectra in the range of 0.5--15 keV yields unacceptable
results due to the fact that (1) the PL component converges to a very steep
index ($\Gamma \sim 3.7$) and dominates the softer component while the STEMS3D
contributes most hard X-ray emissions, (2) $kT$ hits the hard upper limit (0.5
keV), and (3) the obtained column density ($(1.56\pm0.05) \times 10^{22}$
cm$^{-2}$) is significantly higher than the value ($(1.17\pm0.02) \times
10^{22}$ cm$^{-2}$) derived from the single STEMS3D model. By fitting the {\it
NuSTAR}+{\it XMM-Newton} spectra below 12 keV with the STEMS3D model only, we
obtain model parameters (Table \ref{fits}) that are consistent with what have
been obtained from {\it XMM-Newton} spectrum modeled in 0.5--7 keV, as
presented in Paper I.

\begin{deluxetable*}{cccccccccc}
\tabletypesize{\tiny} \tablewidth{0pt} \tablecaption{Spectral Fit Results of
\axpu0142,  \ce1841, \de2259, and \ae1048} \tablehead{\colhead{Source} &
\colhead{nH} & \colhead{$kT$} & \colhead{$B$} & \colhead{$\Delta\phi$} &
\colhead{$\beta$} & \colhead{$\Gamma$}& \colhead{$F_{\rm S}$} &
\colhead{$F_{\rm P}$}
& \colhead{$\chi^2$/dof} \\
\colhead{}  & \colhead{($10^{22}$ cm$^{-2}$)} & \colhead{(keV)} &
\colhead{($10^{14}$ G)} & \colhead{(rad)} & \colhead{} & \colhead{}
 & \colhead{} & \colhead{} &\colhead{}} \startdata

\hline
\axpu0142  &  $0.97_{-0.04}^{+0.04}$ & $0.31_{-0.02}^{+0.02}$ & $6.14_{-0.32}^{+0.32}$ & $1.70_{-0.04}^{+0.02}$ & $0.21_{-0.01}^{+0.01}$ & $0.57_{-0.05}^{+0.04}$ &18.6 & 3.8 & 1290.3/1235\\
\hline
\ce1841    &  $3.36_{-0.31}^{+0.30}$ & $0.35_{     }^{+0.05}$ & $3.99_{-0.27}^{+0.31}$ & $0.83_{-0.12}^{+0.21}$ & $0.28_{-0.02}^{+0.03}$ & $1.23_{-0.05}^{+0.05}$ &3.6 & 5.1 & 1183.1/1162\\
\hline
\de2259  &  $1.17_{-0.09}^{+0.08}$ & $0.33_{-0.02}^{+0.03}$ & $5.60_{-0.36}^{+0.40}$ & $1.60_{-0.16}^{+0.16}$ & $0.20_{-0.01}^{+0.01}$ & $0.20_{-0.23}^{+0.28}$ & 5.0 & 0.58 & 567.6/548\\
\hline
\ae1048$\dag$  &  $1.17_{-0.02}^{+0.02}$ & $0.40_{-0.01}^{+0.01}$ & $2.40_{-0.05}^{+0.05}$ & $1.12_{-0.05}^{+0.04}$ & $0.18_{-0.01}^{+0.01}$ & \nodata &1.4 & \nodata & 1370.0/995
\enddata

\tablecomments{$F_{\rm S}$: unabsorbed flux of the STEMS3D component in 1--79
keV in units of 10$^{-11}$ erg s$^{-1}$ cm$^{-2}$ with cross-normalization
factors taken into account. $F_{\rm P}$: unabsorbed flux of the PL component in
1--79 keV in units of 10$^{-11}$ erg s$^{-1}$ cm$^{-2}$. All errors are in the
90\% confidence level. $\dag$: Spectra are fit with a single STEMS3D
component.\label{fits}}
\end{deluxetable*}

\section{Discussion and Conclusions}

In the framework of the magnetar model, bright X-ray emission is expected to be
primarily fed by a magnetic energy budget through various channels
\citep[e.g.,][]{thompson02, lyutikov06, vigano13}. Therefore, studying
broadband observations of magnetars should offer essential information about
their radiation mechanisms. However, there is still a shortage of physically
oriented magnetar emission models covering both soft and hard X-rays. As a part
of our ongoing effort to build a physically motivated 3D magnetar emission
model, STEMS3D, we extended the model to accommodate spectral investigations up
to 100 keV. We obtain a perfect fit for all of the applied spectra with the
STEMS3D model appended with a PL component (Table \ref{fits}). The model
parameters of STEMS3D are in agreement with our previous investigations of soft
X-ray spectra only (Paper I). (1) The magnetic field strengths $B$ obtained
using different data sets are almost the same, implying that the application of
the STEMS3D model is indeed robust. The magnetic field of \de2259 obtained here
($B=5.6\times10^{14}$ G) is about one order of magnitude larger than the
spin-down inferred value, likely because the assumption of a fixed angle
between the rotation axis and dipole moment in the framework of magnetic dipole
radiation may not be valid in this source. (2) The magnetospheres in these four
sources are strongly twisted ($\Delta\phi \gtrsim 1$). (3) The electron
velocity is non-relativistic ($\beta \lesssim 0.3$). Practically, this means
that soft X-ray photons from the NS surface cannot act as the seed for the
entire hard X-ray emission, and an additional non-thermal process has to
account for the bright hard X-ray radiation of \axpu0142, \ce1841, and \de2259.

The hard X-ray component in magnetars is widely considered to be generated in
the magnetosphere \citep[e.g.,][]{baring07,beloborodov07}. Recently,
\cite{beloborodov13} proposed a relativistic outflow model for the hard X-ray
emission and predicted that an averaged photon index of 1.5 for the hard X-ray
spectrum. However, the observed spectra are always harder than the predicted
value, especially for \axpu0142 and \de2259, which could be due to the effect
of the object inclination to the line of sight. Currently, this model provides
satisfactory fits to the phase-averaged and phase-resolved hard X-ray emission
\citep{an13,hascoet14,vogel14,tendulkar15}. On the other hand, the soft X-ray
spectra of \axpu0142, \ce1841, and \de2259 can be best fit by a Comptonized BB
model or by the superposition of a hotter modified BB and a cool BB.
\cite{beloborodov13} argued that the fitted modified BB corresponding to a hot
spot on the NS surface can be interpreted as the footprint of outflows having
either a polar-cap or a ring shape \citep[e.g., ][]{vogel14}. It is important
to note that this model does not self-consistently calculate the surface
temperature, and only offers phenomenological fits rather than physical fits.
Alternatively, the flux relaxation of the outbursts accompanied by decreasing
NS surface temperature exhibited in some transient magnetars could be
interpreted as the evidence of the crustal cooling \citep[see e.g.,][]{guver07,
scholz14}. By analyzing the multiple observations of variable magnetars, we
suggested that outbursts in SGRs/AXPs could also be driven by different
mechanisms (Paper I).

Our spectral investigations reveal that the STEMS3D (soft band) and the PL
(hard band) components have the comparable flux measurements in the 5--15 keV
energy range (Figure \ref{spec}).  We also find that the fluence of the hard
spectral component starts to dominate the soft component at $\sim$ 10.1, 5.7,
and 13.3 keV for \axpu0142, \ce1841, and \de2259, respectively. The {\it
NuSTAR} spectra of \axpu0142, \ce1841, and \de2259 above 15 keV are represented
by a PL model of the photon indices $0.65_{-0.05}^{+0.06}$,
$1.23_{-0.05}^{+0.05}$, and $0.25_{-0.24}^{+0.25}$, respectively. Although, for
three sources only, we find an anti-correlation between the photon index and
the flux domination energy, as may be naturally expected from a PL dependence.

The PL component of \axpu0142 obtained from {\it NuSTAR} is harder than those
measured from {\it INTEGRAL} \citep[$\sim 0.93 \pm 0.06$][]{den08b} and {\it
SUZAKU} \citep[$\Gamma \sim 0.89^{+0.11}_{-0.10}$][]{enoto11}. Investigating
{\it INTEGRAL} data spanning 2003--2011, \cite{wang14} reported the variation
of the hard X-ray spectral shape. The hard X-ray component can be fit with a
simple PL model and the photon index changes from 0.7 to 1.6; meanwhile, a
cutoff PL model provides even better fits to the data with the photon index
varying between 0.3 and 1.5 and the cutoff energy in the range $\sim 110-250$
keV. It is suggestive that this source is mostly in a hard X-ray spectral state
of $\Gamma \sim 0.9$ and makes intermittent transitions to a spectrally harder
state, and {\it NuSTAR} observations were likely performed during the spectral
transition.

Investigating {\it Suzaku} data for nine magnetars, \cite{enoto10} found a
tight anti-correlation between the characteristic age inferred from spin-down
rates and the hardness ratio (i.e., the flux ratio of hard to soft spectrum
component in 1--60 keV). In other words, the hard X-ray component decreases
with NS age. We calculate the 1--79 fluxes of the STEMS3D and PL components
(Table \ref{fits}), and find that the data for \axpu0142, \ce1841, and \de2259
are in agreement with this relation. Surprisingly, there is no significant hard
X-ray emission observed in \ae1048, which is the youngest in characteristic age
($\sim 4.5$ kyr) among these four sources. If we assume that \ae1048 would have
the same spectral hardness as \ce1841 (with a characteristic age of $\sim 4.6$
kyr), then we would expect a flux level of $\sim 2.1\times10^{-11}$ erg
s$^{-1}$ cm$^{-2}$ in 10--79 keV, which is well above the {\it NuSTAR}
detection limit, indicating different spectral hardness values in these two
sources. Note that the background components of {\it NuSTAR} observations vary
from detector to detector, and it might be inappropriate to extract background
emission from any arbitrary source-free region in the field of view
\citep{wik14}. Generally, the choice of background does not affect the analysis
of the bright source \citep[e.g., \de2259,][]{vogel14}. In order to check any
influence from background selection, we also generated the background spectra
using the tool \texttt{nuskybgd} developed by \cite{wik14}, and obtained nearly
identical modeling results. As an example, we plot the source as well as the
background spectra extracted from both the canonical analysis procedure and the
background model \texttt{nuskybgd} for the longest observation (ObsID =
30001024005)
\footnote{{\protect\url{https://github.com/NuSTAR/nustar-idl/tree/master/nuskybgd}}}
in Figure \ref{bkg}. The background spectra produced in different ways
marginally deviate from each other at soft X-rays ($\lesssim$15 keV) where the
source emission dominates the background. Beyond 20 keV, the background spectra
are consistent with each other, which is a significant fraction of the source
photons. The non-detection of persistent hard X-ray emission suggests that the
additional mechanism responsible for the hard X-ray emission is not functioning
efficiently in \ae1048, as it does in other persistent sources.

\acknowledgments{We would like to thank the referee for helpful
suggestions and comments. We thank Feryal \"Ozel for providing the highly
magnetized NS surface emission code. S.S.W. is supported by the Scientific and
Technological Research Council of Turkey (T\"{U}B\.{I}TAK) and EC-FP7 Marie
Curie Actions-People-COFUND Brain Circulation Scheme (2236).}

\end{document}